\documentclass[bibyear]{aa}
\usepackage{float}
\usepackage{color}
\usepackage{graphicx}
\usepackage{ulem}
\usepackage{txfonts}
\date{}

\usepackage[switch]{lineno}

\usepackage{lineno}
\begin{document} 

   \title{Hurricanes on tidally locked terrestrial planets: Fixed surface temperature experiments}

   \author{Mingyu Yan
          \inst{1},
          \and
          Jun Yang \inst{1,*}}

   \institute{\inst{1}Department of Atmospheric and Oceanic Sciences, School of Physics, Peking University, Beijing 100871, China.\\
   \inst{*}Corresponding author: Jun Yang: junyang@pku.edu.cn }

  \abstract
{}{In this work, we study the presence of hurricanes on exoplanets. Tidally locked terrestrial planets around M dwarfs are the main targets of space missions looking to discover habitable exoplanets. The question of whether hurricanes can form on this kind of planet is important for determining their climate and habitability.}{Using a high-resolution global atmospheric circulation model, we investigated whether there are hurricanes on tidally locked terrestrial planets under 
fixed surface temperatures (T$_S$). The relevant effects of the planetary rotation rate, surface temperature, and bulk atmospheric compositions were examined.}{We find that hurricanes can form on the planets but not on all of them. For planets near the inner edge of the habitable zone of late M dwarfs, there are more numerous and stronger hurricanes on both day and night sides. For planets in the middle and outer ranges of the habitable zone, the possibility of hurricane formation is low or even close to zero, as has been suggested in recent studies. Earth-based hurricane theories  are applicable to tidally locked planets only when the atmospheric compositions are similar to that of Earth. However, if the background atmosphere is lighter than H$_2$O, hurricanes can hardly be produced because convection is always inhibited due to the effect of the mean molecular weight, similarly to the case of Saturn. These results have broad implications on the precipitation, ocean mixing, climate, and atmospheric characterization of tidally locked planets. Finally, A test with a coupled slab ocean and an Earth-like atmosphere in a tide-locked orbit of ten Earth days demonstrates that there are also hurricanes present in the experiment.}{}

\authorrunning{Yan \& Yang}
\titlerunning{Hurricanes on tidally locked planets}
   \maketitle

\section{Introduction}

Hurricanes (also known as tropical cyclones or typhoons) are low-pressure weather systems with well-organized convection, which are among some of the most destructive disasters on Earth because they can induce strong winds, a rapid rise in local sea level, and heavy precipitation (Anthes \cite{Anthes}, Emanuel \cite{Emanuelb}). Hurricanes can also enhance the vertical mixing of heat and nutrients in the ocean, increase horizontal oceanic heat transport, and, subsequently, influence global climate (Emanuel \cite{Emanuela}, Jansen \&  Ferrari \cite{Jansen}, Li \& Sriver \cite{LiH}). For instance, the curl of strong hurricane winds can cause divergence and convergence in the upper ocean, producing regions of up-welling and down-welling, enhancing the exchanges between surface and subsurface oceans. Therefore, it is an important and interesting issue to consider whether hurricanes can form on other potentially habitable planets beyond Earth.

In this study, we focus on tidally locked terrestrial planets around M dwarfs due to their relatively large planet-to-star ratios and frequent transits based on observations. These planets differ from Earth in three major aspects: the uneven distribution of stellar energy between the permanent day and night sides, the slow rotation rate due to strong tidal force, and the redder stellar spectrum than the Sun. Several atmospheric general circulation models (AGCMs) have been employed and modified to simulate and understand the atmospheric and climatic dynamics of the planets  (Joshi et al. \cite{Joshi}, Merlis \& Schneider \cite{Merlis}, Edson et al. \cite{Edson}, Pierrehumbert \cite{Pierrehumberta}, Wordsworth et al. \cite{Wordsworthetal}, Leconte et al. \cite{Lecontea}, Menou \cite{Menou}, Showman et al. \cite{Showmanb}, Carone et al. \cite{Carone}, Wordsworth \cite{Wordsworth}, Shields et al. \cite{Shields}, Turbet et al. \cite{Turbetetal}, Boutle et al. \cite{Boutle}, Checlair et al. \cite{Checlair}, Haqq-Misra et al. \cite{Haqq-Misra}, Kopparapu et al. \cite{Kopparapu}, Noda et al. \cite{Noda}, Wolf \cite{Wolf2017}, Turbet et al. \cite{Turbet}, Del Genio et al. \cite{Del Genio}, Pierrehumbert \& Hammond \cite{Pierrehumbertb}, Yang et al. \cite{Yang}). 

These AGCMs have horizontal resolutions that are always equal to or larger than 300 km, so that most of their studies focus on planetary-scale phenomena, such as global-scale Walker circulation, equatorial superrotation, and forced Rossby and Kelvin waves, and their models are not able to properly simulate the characteristic features of hurricanes, such as the warm core, the eye-eyewall structure, and the spiral rain bands. No work has investigated synoptic phenomena apart from Bin et al. (\cite{Bin}). Based on the output data of an AGCM, the authors estimated genesis potential index of hurricanes and showed that the probability of hurricane formation is low for planets in the middle range of the habitable zone of M dwarfs. However, the model resolution they used was not directly capable of simulating hurricanes and the question of whether the empirical index could be applied to exoplanets could not be answered. Moreover, they considered planets in the middle range of the habitable zone only. Here, we show that the possibility of hurricane formation increases with temperature and for planets with higher temperatures (closer to the inner edge of the habitable zone), the possibility is greater.

In this study, we explicitly simulate hurricane formation on tidally locked terrestrial planets with a high-resolution ($\approx$50~km) AGCM. The structure of the paper is as follows. Section~2 describes our methods, Section~3 presents our results, and Section~4 gives our conclusions.


\section{Model description and experimental design}
For our model, we used the global Community Atmosphere Model version 4 (CAM4) with a dynamical core of finite volume (Neale et al. \cite{Neale}). Deep convection was parameterized using the updated mass flux scheme of Zhang and McFarlane (\cite{ZhangG}). Subgrid-scale momentum transport associated with convection was included (Richter \& Rasch \cite{Richter}). The parameterization of shallow moist convection is based on Hack (\cite{Hack}). Condensation, evaporation, and precipitation parameterization is based on Zhang et al. (\cite{ZhangM}) and Rasch and Kristjansson (\cite{Rasch}). Cloud fraction is diagnosed from atmospheric stratification, convective mass flux, and relative humidity (Slingo \cite{Slingo}, Hack et al. \cite{Hack1993}, Kiehl et al. \cite{Kiehl}, Rasch \& Kristjansson \cite{Rasch}). The realistic radiative transfer of water vapor, clouds, greenhouse gases, and aerosols are included as well (Ramanathan \& Downey \cite{Ramanathan}, Briegleb \cite{Briegleb}, Collins et al. \cite{Collins}, Neale et al. \cite{Neale} ).

The horizontal resolution we employed is 0.47$^{\circ}$\,$\times$\,0.63$^{\circ}$ in latitude and longitude, respectively. The number of vertical levels is 26. The planetary surface is covered by seawater throughout (namley, an aquaplanet). Because of the high resolution and limited computational power, we specify surface temperature (T$_S$) in the simulations. With a fixed T$_S$, the atmosphere reaches an equilibrium state within several years. If the model were coupled to a 50-m slab ocean, the surface and atmosphere would require tens of years to reach the equilibrium state, which is about one order of magnitude longer than that in the simulations with fixed T$_S$. The thermal inertia of the slab ocean is much larger than that of the atmosphere. If the model were coupled to a fully dynamical ocean with a depth of, for example, 3000 m, the model will require thousands of years to reach the equilibrium state due to the high thermal inertia and the slow motion of the deep ocean. For simulating hurricanes, a fixed surface temperature experiment is a good start and a useful method for understanding the formation and the properties of hurricanes, as found in hurricane simulations on Earth, carried out, for example, in the studies of Held and Zhao (\cite{Held}) and Khairoutdinov and Emanuel (\cite{Khairoutdinov}) and the recent review papers of Emanuel (\cite{Emanuelb}) and Merlis and Held (\cite{Merlis19}). The fixed-temperature surface acts as a boundary condition for the atmosphere system. Under a fixed T$_S$, the surface and the top of the atmosphere are not in energy balance, but the atmosphere itself is in energy balance; this is because the energy deficit or excess at the surface is approximately equal to that at the top of the atmosphere. In the coupled slab ocean experiment (shown in Sect. 4 below), the surface, the top of the atmosphere, and the atmosphere are all in energy balance.

The surface temperature is set according to previous simulations of lower resolution AGCMs coupled to a 50-m slab ocean (Yang et al. \cite{Yang13}, Wolf et al. \cite{Wolf}). On the day side, the surface temperature is a function of latitude and longitude: $(T_{max}-T_{min})cos(\varphi)cos(\lambda)+T_{min}$, or $(T_{max}-T_{min}){cos}^{1/4}(\varphi){cos}^{1/4}(\lambda)+T_{min}$, where $T_{max}$ is the maximum surface temperature, $T_{min}$ is the minimum surface temperature, $\varphi$ is the latitude, and $\lambda$ is the longitude. On the night side, the surface temperature is uniform with a value of $T_{min}$. Three groups of $T_{max}$ and $T_{min}$ are used. One is for planets near the inner edge of the habitable zone, 315 K \& 310 K. The other two are for planets in the middle range of the habitable zone, 308 K \& 275 K and 301 K \& 268 K; the power of 1/4 is used due to the very weak temperature gradients in the substellar region and strong temperature gradients near the terminators.

The planetary rotation period is set equal to the orbital period. Four rotation periods are examined: 6, 10, 20, and 40 Earth days. For other types of spin-orbit resonance, such as 3:2 as in the case of Mercury, the climate lies between the synchronous rotation and the rapid rotation of Earth (Yang et al. \cite{Yang14}) and we have not carried out these kinds of experiments to date. Planetary radius and gravity are set to be the same as Earth, but both obliquity and eccentricity are set to zero. Stellar temperature is set to 2,600 or 3,700 K. The stellar radiation at the substellar point is set to 1,300 or 1,800 W m$^{-2}$. By default, the mean surface pressure is 1.0 bar with $\approx$79\% N$_2$ and $\approx$21\% O$_2$. For greenhouse gases, we set the CO$_2$ concentration to 367 parts per million by volume (ppmv), N$_2$O to 316 parts per billion by volume (ppbv), and CH$_4$ to 1760 ppbv. The ozone concentration is set to be the same as present-day Earth, which may influence the outflow temperature of hurricanes and the overshooting of extremely strong convection.

In order to briefly test the effect of atmospheric composition on hurricane formation, we did several ideal experiments in which the background gas is set to H$_2$, He, N$_2$, O$_2$, and CO$_2$, respectively. The corresponding mean molecular weights are 2.02, 4.00, 28.01, 31.99, and 44.00 g mole$^{-1}$ and the corresponding specific heats (Zhang \& Showman \cite{ZhangX}) are 28.9, 20.8, 29.1, 29.5, and 37.2 J mole$^{-1}$ K$^{-1}$. We modify these two constants only. The model we employed is incapable of calculating the radiative transfer of dense H$_2$, He, O$_2$, or CO$_2;$  meanwhile, surface temperatures under background gases that differ from Earth have not been seriously examined. Thus, we chose to use the globally uniform surface temperature (301 K) and stellar radiation (340 W m$^{-2}$) with neither a seasonal nor diurnal cycle. This idealized thermal boundary condition is unrealistic, but it is capable of avoiding the effects of any strong wind shear, the baroclinic zone, or other features that may inhibit hurricane formation or propagation (Merlis \& Held \cite{Merlis19}). We used two planetary rotation periods:\ of one and of three Earth days.

The initial states of the experiments were based on long-term (of 40 Earth years) simulations using a lower resolution of 4$^{\circ}$$\times$5$^{\circ}$ or 1.9$^{\circ}$$\times$2.5$^{\circ}$ under the same experimental designs and parameterization schemes. Then each experiment was run for five Earth years under high resolution and the last four years were used to carry out the analysis, presented below.

Hurricane formation and tracking is based on six hourly model output variables using the Geophysical Fluid Dynamics Laboratory tracking algorithm (Zhao et al. \cite{Zhao}). Candidate hurricanes are identified by finding regions that satisfy the following criteria: 1) the local 850-hPa relative vorticity maximum exceeds 3.5$\times$10$^{-5}$ s$^{-1}$; 2) the 850-hPa warm-core temperature must be at least 0.5 K warmer than the surrounding local mean; 3) the distance between the local sea level pressure minimum and the vorticity maximum should be within a distance of 2$^{\circ}$ latitude or longitude and so, this should also be the distance between the local sea level pressure minimum and the warm-core center; 4) the maximum 850-hPa wind speed exceeds $\approx$33 m s$^{-1}$ at some point. These values for the thresholds impact the exact number of detected hurricanes but they do not affect the main conclusions of this study.

\begin{figure*}
\centering
\setlength{\abovecaptionskip}{0.1cm}
\includegraphics[width=0.9\textwidth]{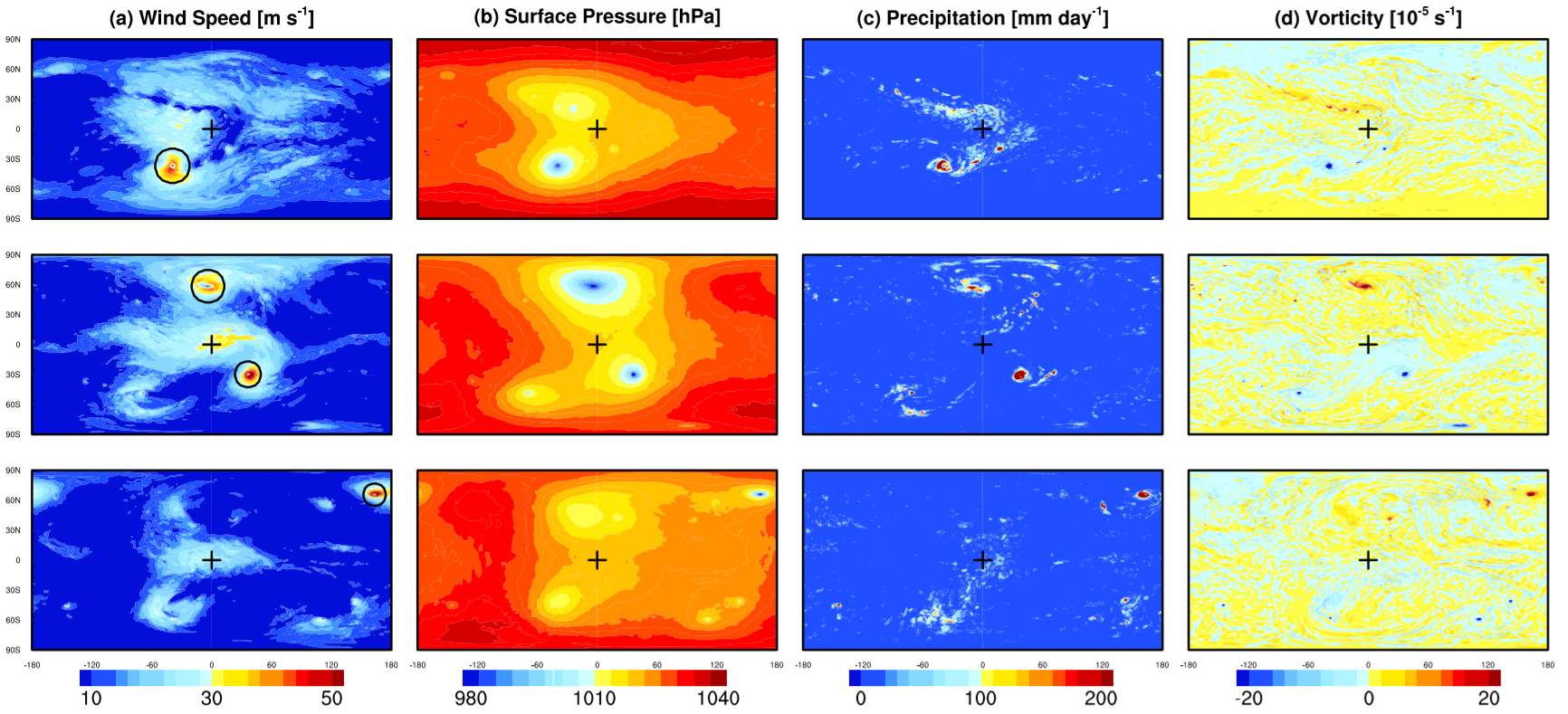}
\caption{Snapshots of hurricanes on a tidally locked aqua-planet near the inner edge of the habitable zone in the control experiment. From left to right, the variables are: instantaneous wind speed at 850 hPa, surface air pressure, precipitation, and the vertical component of relative vorticity at 850 hPa, respectively. From upper to bottom, they are for three different moments. The four hurricanes are marked with black circles over the wind speed panels. The black cross is the substellar point in this figure and hereafter. See the supplementary video online for a visualisation of the evolution of the hurricanes.}
\label{fig1}
\end{figure*}

For Earth, a useful method for estimating the possibility of hurricane formation is the genesis potential index (GPI; Emanuel \& Nolan \cite{Emanuelc}), which is written as:
\begin{equation} 
GPI = |10^5 (\zeta + f)|^{3/2}(RH/50)^3(V_{pot}/70)^3(1.0+0.1V_{shear})^{-2},
\end{equation}
where $\zeta$ is the vertical component of relative vorticity, $f$ is the planetary vorticity, $RH$ is the relative humidity at the middle troposphere (600 hPa), $V_{shear}$ is the wind shear of horizontal winds between the upper and lower troposphere (300 minus 850 hPa; or called vertical wind shear), and $V_{pot}$ is potential intensity. The $V_{pot}$ is a measure of the maximum near-surface wind that can be maintained by hurricane under given environmental conditions. We note that these parameters are not entirely independent; for instance, vertical wind shear can influence relative humidity. The value of $V_{pot}$ is calculated based on a local balance between thermal energy import and mechanical energy dissipation (Emanuel \cite{Emanuel}, Bister \& Emanuel \cite{Bister}), written as

\begin{equation}
    V_{pot}^2 = \frac{C_k}{C_D}\frac{T_s}{T_o}\,[CAPE^* - CAPE^b]|_m,
\end{equation}
where $C_k$ is the exchange coefficient for enthalpy, $C_D$ is the drag coefficient, $T_s$ is the surface temperature, $T_o$ is the mean outflow temperature, ${CAPE}^\ast$ is the convective available potential energy of air lifted from saturation at sea surface, and ${CAPE}^b$ is that of the boundary layer air. Both ${CAPE}^\ast$ and ${CAPE}^b$ are computed at the radius of maximum surface wind.

In the discussion of the size of hurricanes, the Rossby deformation radius ($L_R$) is used, as described in Section 3.3 below. Here, $L_R$ is the length scale at which rotational effects become as important as the effects of gravity waves or buoyancy in the evolution of the flow in a disturbance. The $L_R$ is equal to $\frac{NH\ }{f}$, where $N$ is the Brunt-Vaisala frequency, and $f$ is the Coriolis parameter. Furthermore, $H$ is the scale height, equaling to $\frac{R^\ast\bar{T}}{M_dg}$, where $R^\ast$ is the universal gas constant, $\bar{T}$ is the mean air temperature, $M_d$ is the molar weight of the atmosphere, and $g$ is the gravity (Wallace \& Hobbs \cite{Wallace}). The $L_R$ decreases as $M_d$ increases due to the reduction of $H$. In idealized experiments with uniform surface temperature or uniform rotation, it is one of the rough scales that can be used for understanding hurricane size (Held \& Zhao \cite{Held}); however, in more realistic conditions such as on Earth, $L_R$ is not a good scaling (Chavas et al. \cite{Chavas16}, Chavas \& Reed \cite{Chavas19}).


\begin{figure*}
\centering
\setlength{\abovecaptionskip}{0.1cm}
\includegraphics[width=0.8\textwidth]{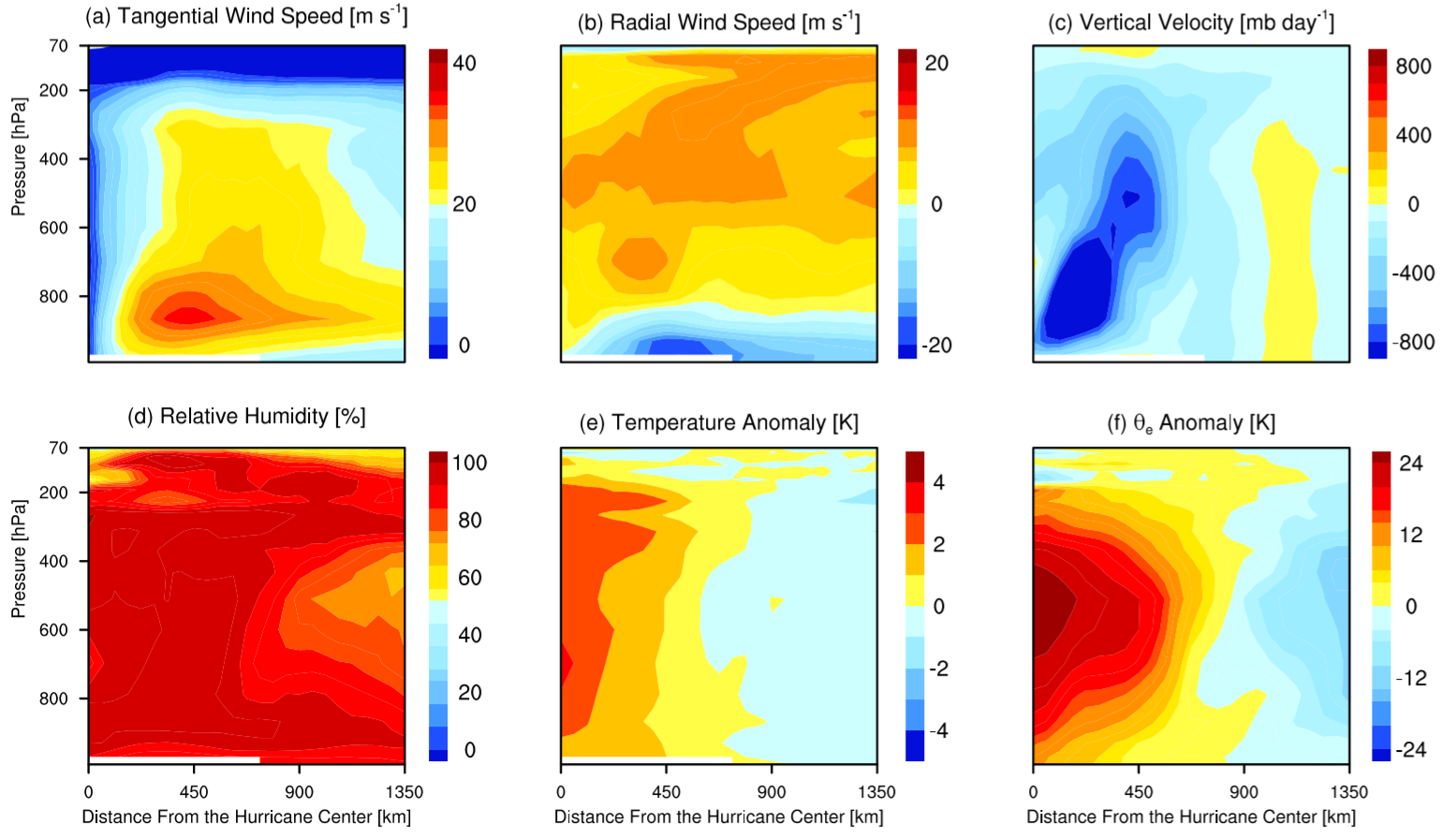}
\caption{Azimuthal-mean height-radius cross-section of a typical, mature hurricane on the day side of a tidally locked aqua-planet in the control experiment. (a) tangential wind speed ($v_\theta$), (b) radial wind speed ($v_r$), (c) vertical velocity ($\omega$), (d) relative humidity, (e) temperature anomaly from the environmental value ($\Delta$T), and (f) equivalent potential temperature anomaly ($\Delta \theta_e$).}
\label{fig2}
\end{figure*}

\begin{figure*}
   \centering
   \includegraphics[width=0.8\textwidth]{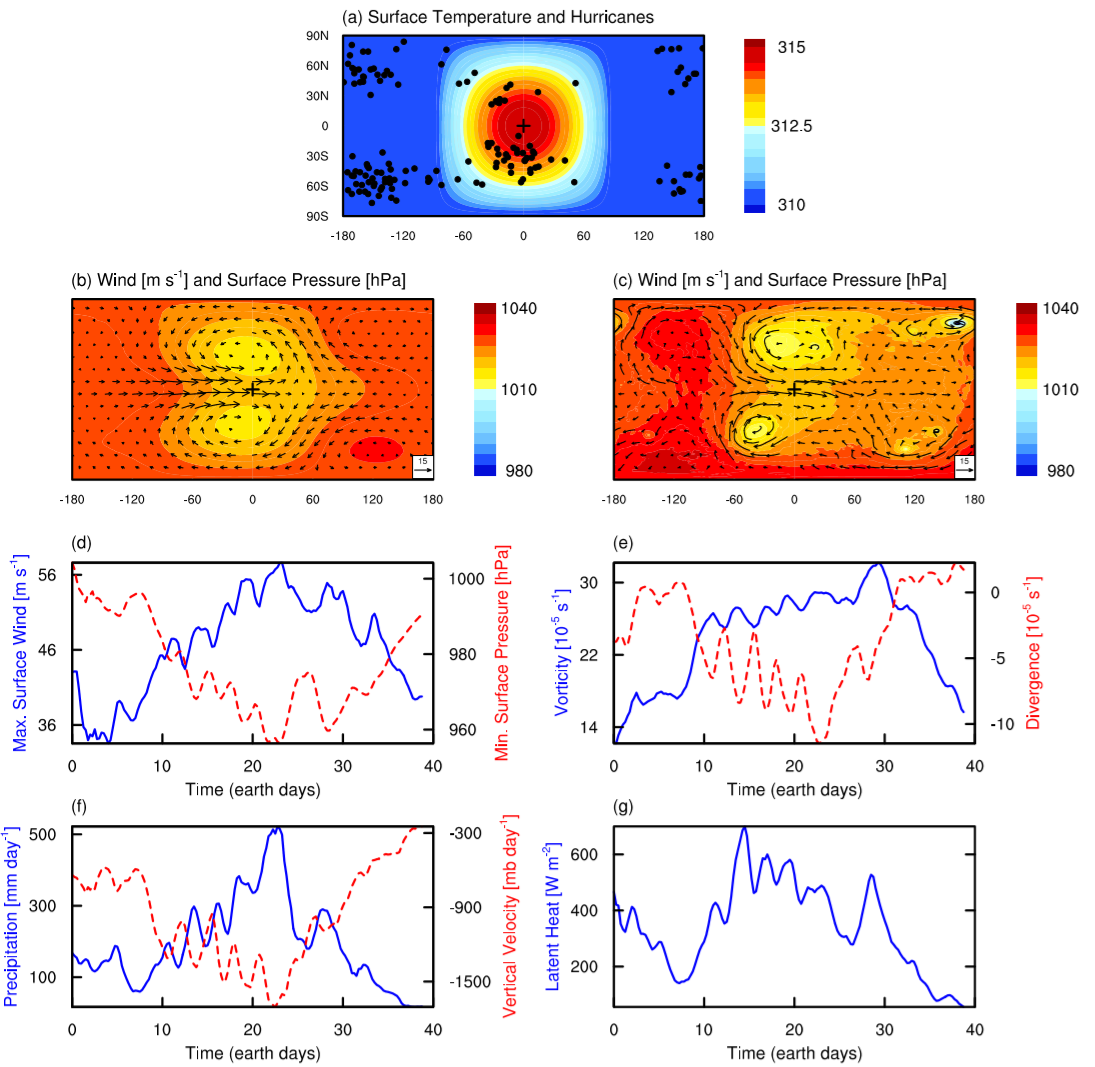}
      \caption{Mechanisms for hurricane formation in the control experiment. (a): Location of hurricane formation (dots) and surface air temperature (color shading). The number of hurricanes during the four-year integration is 154. (b): Long-term mean surface air pressure (shading) and winds at 850 hPa (vector). (c): Same as (b) but for an instantaneous. The life cycle of one hurricane on the day side: (d): Maximum surface wind speed (blue) and minimum surface pressure (red), (e): Relative vorticity (blue) and divergence (red) at 850 hPa, (f): Precipitation (blue) and vertical velocity at 850 hPa (red), and (g): Surface latent heat flux (blue). For (e)-(g), the variables are calculated for area mean of 500$\times$500 km$^{2}$ around the low-pressure center.}
         \label{fig3}
   \end{figure*}

\section{Results}
\subsection{Hurricanes on tidally locked planets}

Figure 1 and 2 show the results of the control experiment for an aquaplanet orbiting close to the inner edge of the habitable zone around a late M dwarf of 2600 K. The rotation period is set to six Earth days. Both the day and night sides are set to hot (310--315~K), and the day-to-night surface temperature contrast is small (see Fig. 3a). This experimental design is due to the fact that the day-to-night atmospheric latent heat transport is very efficient for planets near the inner edge (Haqq-Misra et al. \cite{Haqq-Misra}, Yang et al. \cite{Yang}); oceanic heat transport can act to further reduce the day-to-night temperature contrast (Yang et al. \cite{Yang}). The high surface temperature, weak day-to-night contrast, and relatively fast rotation rate (comparing to planets in the middle range of the habitable zone or planets orbiting around hotter stars) benefit hurricane formation. The temperatures of 310-315 K are close to the conditions of a runaway greenhouse state. For tidally locked planets, the planets start to enter runaway greenhouse state when the maximum surface temperature is close to or higher than these values (Wolf et al. \cite{Wolf}).

Clearly, there are hurricanes in the control experiment. In the mature stage of the hurricanes, maximum wind speed reaches $\approx$30-50 m s$^{-1}$ (Fig.~1a), surface air pressure at the center is $\approx$950-980 hPa (Fig.~1b), precipitation reaches as high as 200-500 mm per day due to strong convection especially near the eyewall (Fig.~1c), and relative vorticity near the surface is on the order of 10$^{-4}$ s$^{-1}$ (Fig.~1d), values that are close to those on Earth (Anthes \cite{Anthes}, Emanuel \cite{Emanuelb}). The surface wind speed increases as the eyewall is approached from outside, but inside the eyewall the winds as well as precipitation weaken rapidly. The winds rotate counter-clockwise in the northern hemisphere and clockwise in the southern hemisphere due to the Coriolis force, although in this experiment, it is much smaller than that on Earth. The precipitation exhibits well-defined spiral bands rather than uniformly distributed throughout the region of the hurricane. The clear patterns of eye-eyewall and spiral rain bands suggest that the model resolution of 50 km is good in resolving the hurricanes.

\begin{figure*}
   \centering
   \includegraphics[width=0.8\textwidth]{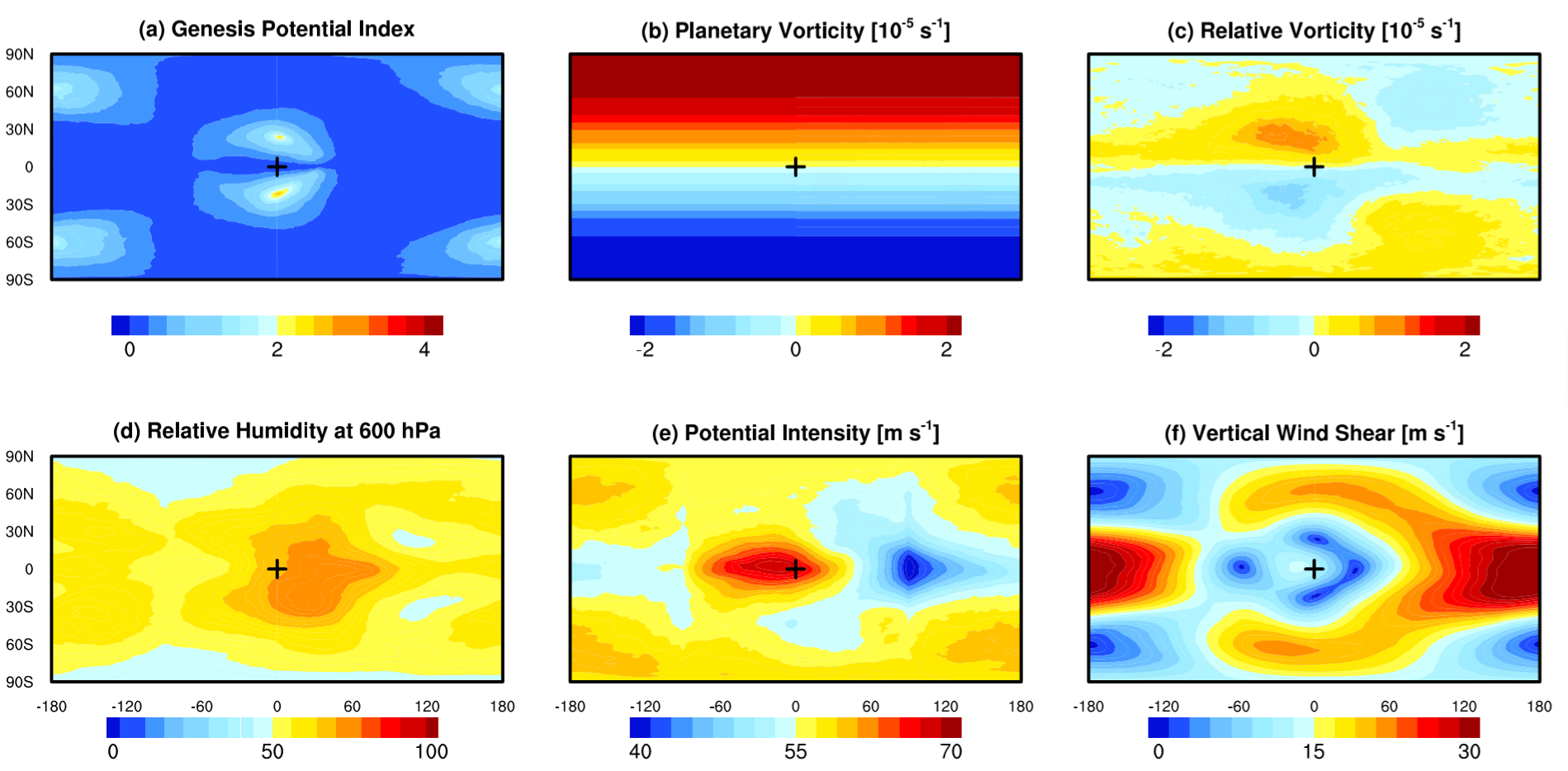}
      \caption{Environmental conditions for hurricane formation. (a): Genesis potential index (GPI); (b): Planetary vorticity; (c): Long-term mean relative vorticity at 850 hPa; (d): Relative humidity at 600 hPa; (e): Potential intensity; (f): Vertical shear of the horizontal winds between 300 and 850 hPa. We note the environmental vorticity in panel (c) is one order smaller than the vorticity of hurricanes shown in Fig.~1d.
              }
         \label{fig4}
   \end{figure*}

For vertical cross structure (Fig.~2), the tangential wind component dominates the flow through the system, although the radial wind is also significant. In the boundary layer, the winds flow towards the low-pressure center. Within the hurricane, upward motion is robust and tilts radially outward. The maximum ascendance near the surface locates at a distance of $\approx$250 km outward from the hurricane center rather than at the center itself. Indeed, weak downward motion takes up the center. Due to the ascendance and deep convection, relative humidity is high in the hurricane. Latent heat release from the convection and adiabatic warming by compression from the subsidence in the eye produces a warmer region of air with temperatures of $\approx$4~K above the environmental value. This can be called a 'warm core,' which is one of the most characteristic features of a hurricane. The warm core can also be viewed from the equivalent potential temperature anomaly. Hurricanes on the night side (Fig.~S1) are smaller in horizontal size compared to those on the day side, $\approx$500 versus 1500 km, but the vertical structures are similar.

Statistical analysis shows that in the control experiment there are four preferred regions for hurricane genesis: the northern and southern tropics of the day side near the substellar point and the middle-to-high latitudes of the night side on each hemisphere (black dots in Fig.~3a). Hurricane formation is largely determined by small-scale convection, large-scale environmental conditions, and the interactions between them. Below, we explore the underlying mechanisms in two ways.

One way is the positive feedback between cumulus convection and larger-scale disturbance, known as the conditional instability of the second kind (CISK; Charney \& Eliassen \cite{Charney}, Smith \cite{Smith}, Yamasaki \cite{Yamasaki}, Wang \cite{Wang}). On tidally locked planets, long-term mean atmospheric circulation is characterized by large-scale Rossby waves on the west and pole of the substellar point and Kelvin waves on the east of the substellar point (Fig.~3b), excited from the uneven stellar radiation distribution (Showman \& Polvani \cite{Showmana}, Showman et al. \cite{Showmanb}). The wave pattern is similar to the tropical Matsuno-Gill pattern on Earth (Matsuno \cite{Matsuno}, Gill \cite{Gill}), but the meridional (south-north) scale is larger, $\approx$10,000 versus 3,000 km. The Rossby waves have one low-pressure center on each hemisphere, whose corresponding environmental vertical motion is updrafts and relative vorticity is positive (negative) on the northern (southern) hemisphere. This low-pressure system favors the onset of the CISK feedback: surface winds spiral into the low-pressure center and create horizontal convergence; this low-level convergence enhances the relative vorticity through vortex stretching, increases upward motion following the conservation of mass, and, in the meantime, brings water vapor into the center, amplifying cumulus convection and release of latent heat. The latent heat release warms the air and lowers the air density through forcing more upper-level air to move outward away from the center, subsequently reducing the surface pressure; the lower surface pressure further enhances the low-level convergence and increases the growth rate of the relative vorticity through vortex stretching (Fig.~3d-f). This feedback is the key in promoting the growth of small-scale disturbances to hurricanes in the background low-pressure regions. It is similar to that on Earth: hurricane generally forms in the monsoon troughs and the confluence zones where the surface pressure is relatively low, collocated with high cyclonic vorticity, convergent surface winds, and divergent winds aloft (Anthes \cite{Anthes}, Emanuel \cite{Emanuelb}). Moreover, during the formation phase, the latent heat flux from the surface to the boundary layer increases strongly (Fig.~3g), which also contributes to intensifying the hurricane through the feedback of wind-induced surface energy exchange (WISHE; Emanuel \cite{Emanuelb}, Wang \cite{Wang}).

 The lifetime of the hurricanes on the day side is $\approx$40-50 Earth days, longer than that on Earth. This is mainly due to the absence of continents and the warm surface everywhere in the experiment. On the night side, the lifetime is shorter, $\approx$10-20 Earth days.

 \begin{figure*}
   \centering
   \includegraphics[width=0.8\textwidth]{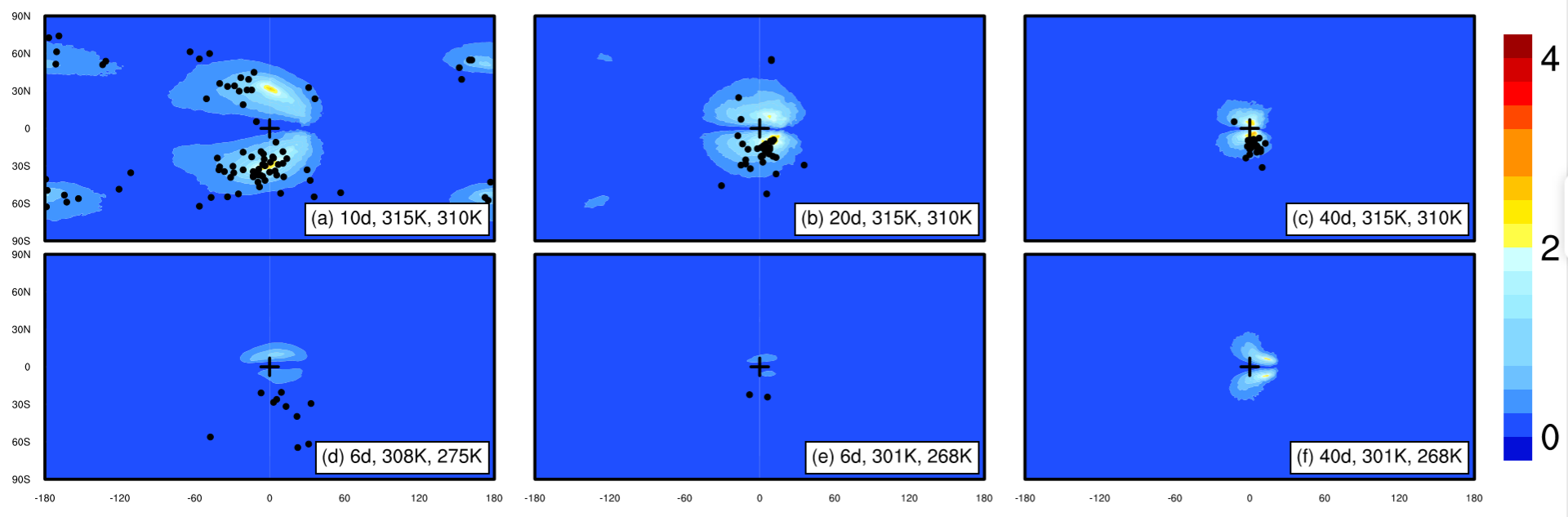}
      \caption{Effects of planetary rotation and surface temperature on hurricane formation (dots) and the GPI (color shading). Experiments are for varying rotation period (a-c), varying surface temperature (d-e), and varying values of both (f). Experimental designs are the same as the control experiment in Fig.~1 except that the rotation period is set to (a): 10, (b): 20, and (c): 40 Earth days; (d): Maximum surface temperature reduced from 315 to 308 K and the night-side surface temperature is reduced from 310 to 275 K (see Fig.~6a); (e): Same as (d) but for 301 K and 268 K, respectively (see Fig.~6b); and (f) same as (e) but for a rotation period of 40 earth days. The number of hurricanes is 88, 34, 21, 10, 2, and 0, respectively. See Fig.~7 for snapshots of typical hurricanes. The southern hemisphere always has more hurricanes than the northern hemisphere; this may be due to some asymmetry in initial state or some stochastic process in the model.
              }
         \label{fig5}
   \end{figure*}

\begin{figure*}
\centering
\setlength{\abovecaptionskip}{0.1cm}
\includegraphics[width=0.8\textwidth]{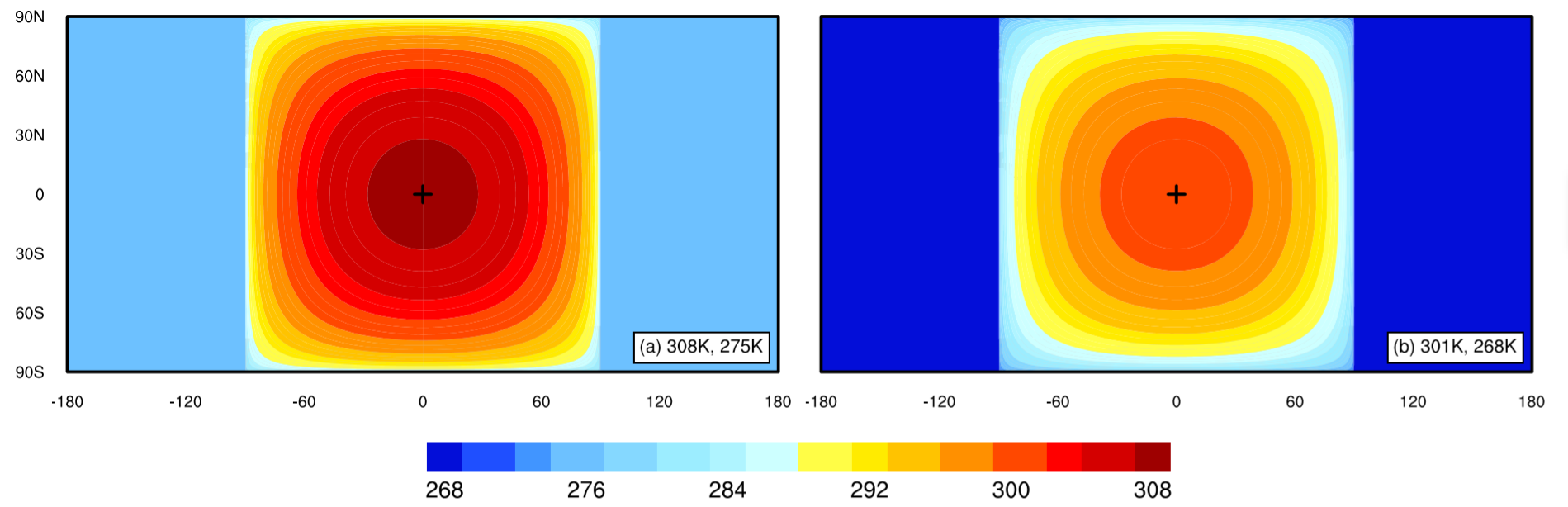}
\caption{Surface air temperatures specified in the simulations of planets in the middle range of the habitable zone. (a): Maximum surface temperature is 308~K and the night-side surface temperature is uniform with a value of 275~K. (b): Same as (a) but the temperature is 7~K lower throughout.}
\label{fig6}
\end{figure*}

\begin{figure*}
\centering
\setlength{\abovecaptionskip}{0.1cm}
\includegraphics[width=0.8\textwidth]{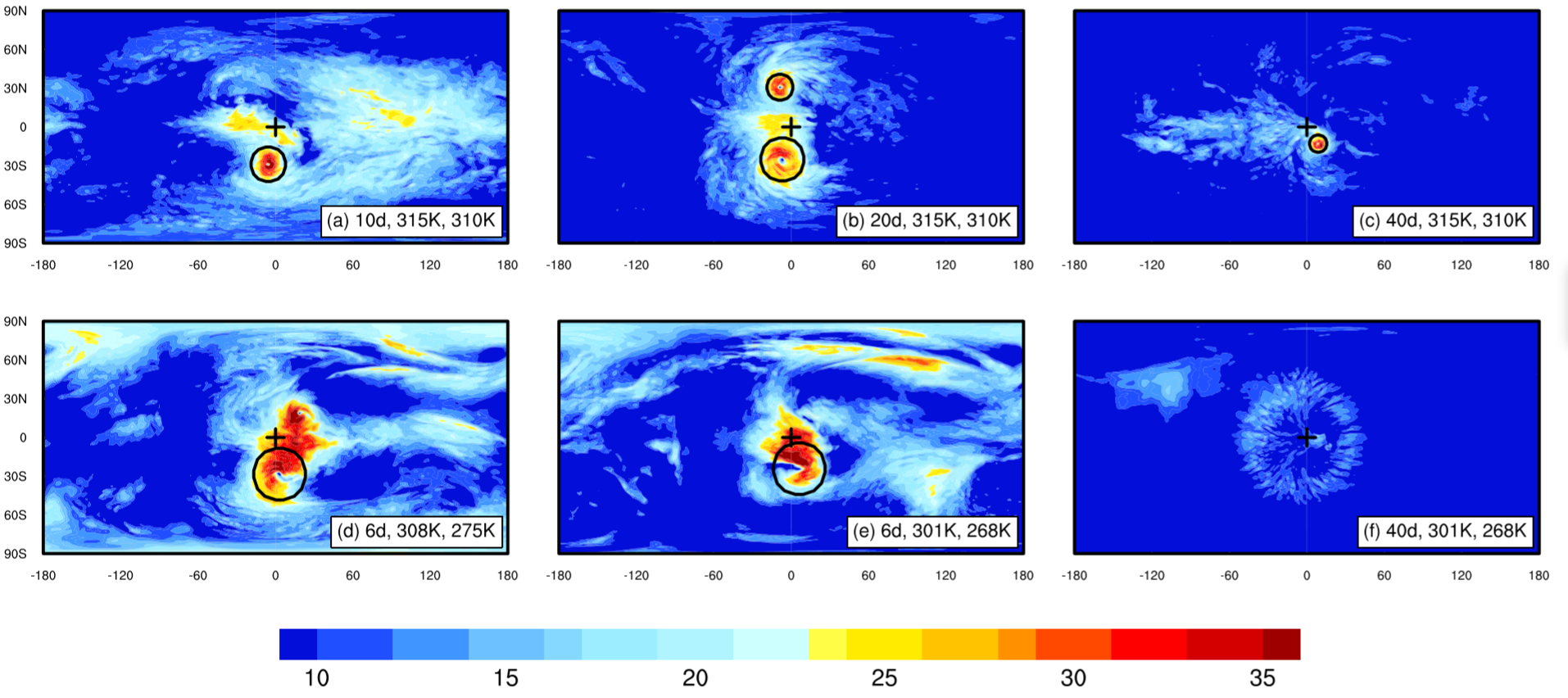}
\caption{Snapshots of instantaneous surface wind speed (m s$^{-1}$) of typical hurricanes in the experiments of varying rotation period (a-c), varying surface temperature (d-e), and variations of both (f). The six hurricanes are marked with black circles. There is no hurricane in panel (f). These experiments are the same as those shown in Fig.~5.}
\label{fig7}
\end{figure*}

Another more quantitative way is the empirical equation of the genesis potential index (GPI; Emanuel \& Nolan \cite{Emanuelc}, Camargo et al. \cite{Camargo}, Bin et al. \cite{Bin}). The index combines five environmental factors to predict the potential of hurricane formation, including planetary vorticity, relative vorticity, relative humidity, potential intensity, and wind shear. A comparison between Fig.~3a and~4a reveals a positive correlation between the location of hurricane genesis and large values of the GPI. In the four hurricane formation regions, GPI values are large because the relative vorticity is great, relative humidity is high over the substellar region, potential intensity is large, and vertical wind shear is weak (Fig.~4b-f). These properties favor hurricane formations in these scenarios. For example, when the shear is strong, an initial disturbance will be ventilated by cooler or drier air and thereby temperature and moisture anomalies are hard to maintain (Tang \& Emanuel \cite{Tang}). In this experiment, the vertical wind shear is strong, especially in the tropics of the night side associated with atmospheric superrotation (Showman et al. \cite{Showmanb}, Pierrehumbert \& Hammond \cite{Pierrehumbertb}) and in the extratropics of the day side (Fig.~4f), so that there is nearly no hurricane formation there. The applicability of the empirical GPI index on Earth to the tidally locked planet is mainly due to the fact that we employed an Earth-like atmosphere here; when the atmospheric composition is quite different from that of Earth, GPI does not serve as a good index, as addressed below. 

The formation of hurricanes on the night side is surprising because the night side has no stellar radiation and the long-term mean vertical motion is downwelling rather than upwelling. In this experiment, based on a planet close to the inner edge of the habitable zone, however, the night-side surface is warm and the surface temperature gradient is small (Fig.~3a). In addition, there are a few short-time, small, low-pressure regions (Fig.~3c), the planetary vorticity is relatively high (Fig.~4b), and the vertical wind shear is weak (Fig.~4f) at the middle-to-high latitudes. These factors promote hurricane genesis there. However, when the surface temperature is decreased or the rotation rate is slowed down, there are fewer or altogether no hurricanes on the night side (see below).

\begin{figure*}
   \centering
   \includegraphics[width=0.8\textwidth]{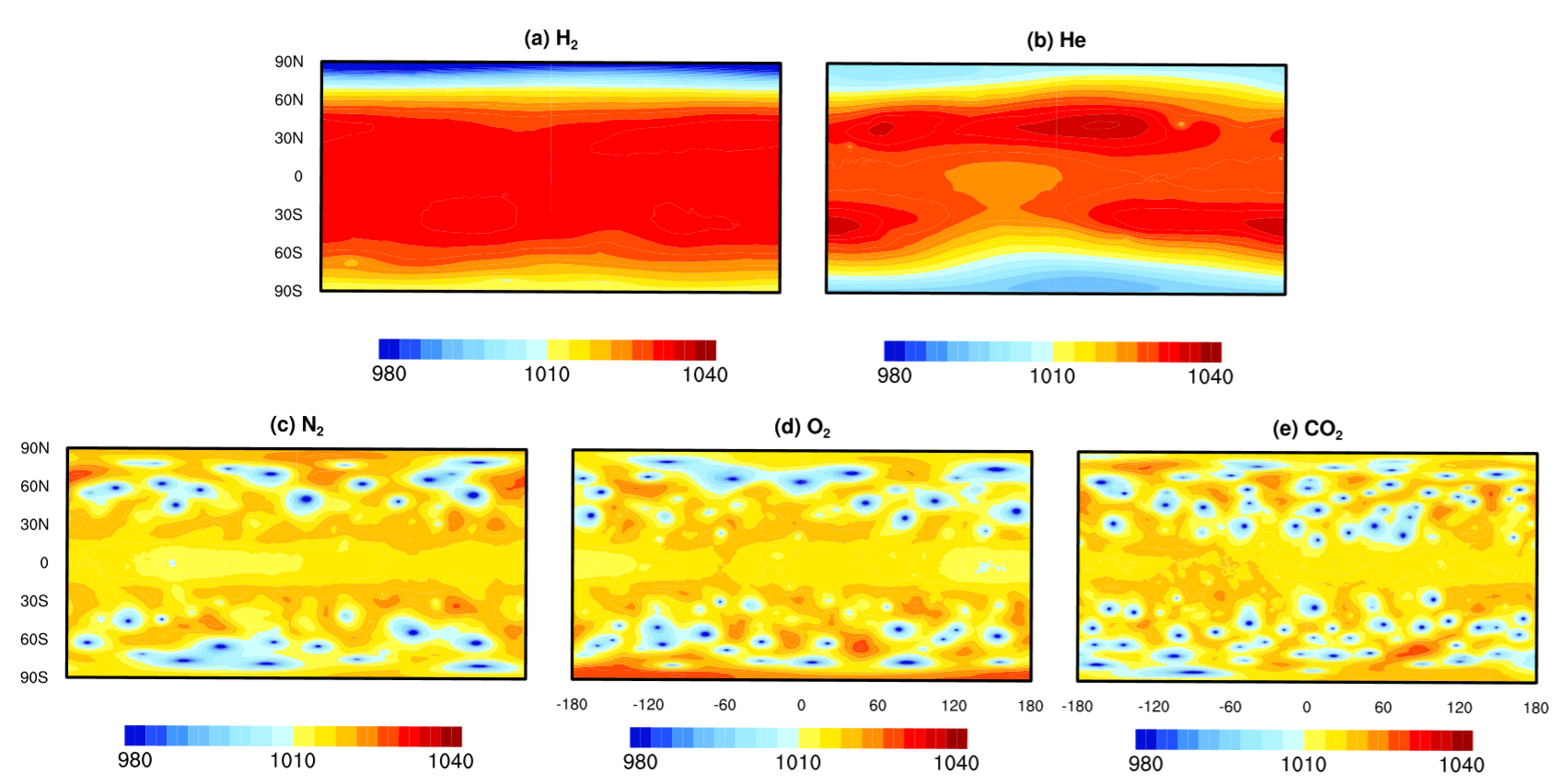}
      \caption{ Effects of background gases on hurricane formation. From (a) to (e), these are snapshots of instantaneous surface air pressure (hPa) under background gases of H$_2$, He, N$_2$, O$_2$, and CO$_2$, respectively. In all these experiments, the planetary rotation period is one Earth day and surface temperature is uniform (301 K). For experiments with a rotation period of three Earth days, the results are the same except that the hurricanes are larger in size but fewer in number in the latter three experiments.
              }
         \label{fig8}
   \end{figure*}

\begin{figure*}
   \centering
   \includegraphics[width=0.8\textwidth]{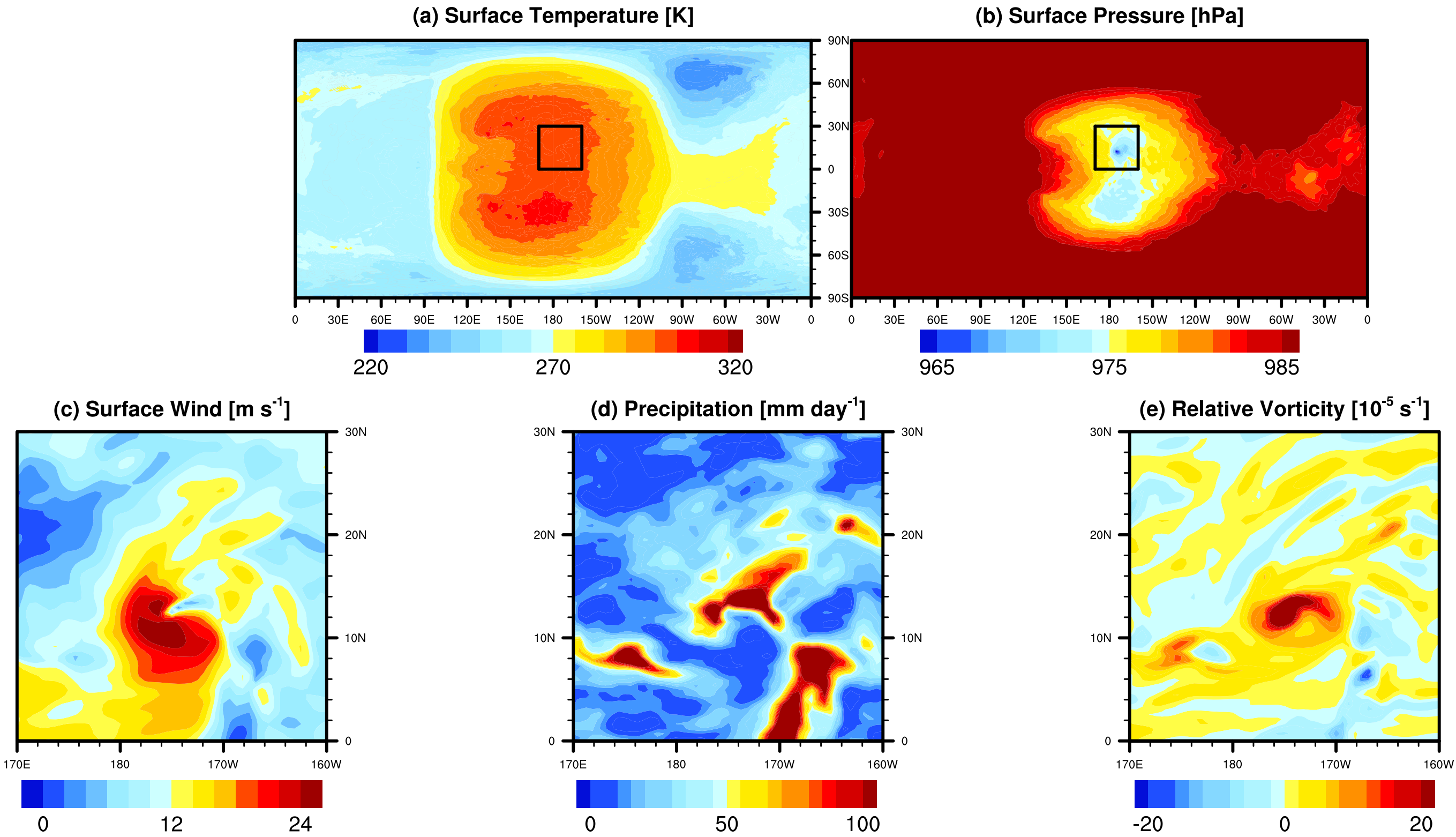}
      \caption{Snapshots of (a) surface temperature, (b) surface air pressure, (c) near-surface wind strength, (d) precipitation, and (e) relative vorticity in one experiment coupled to a slab ocean. Panels c–e: only the hurricane region is shown in order to more clearly exhibit the structure of the hurricane. In this experiment, rotation period (=orbital period) is ten Earth days, the CO$_2$ concentration is 300 ppmv, stellar flux is 1450 W m$^{-2}$, and the star temperature is 2600 K. No oceanic heat transport is involved in this run.      }
         \label{fig9}
   \end{figure*}

\subsection{Effects of planetary rotation rate and surface temperature} 
In order to test the effect of the rotation rate, we performed three experiments in which rotation period is increased (i.e., the rotation rate is decreased) while other experimental design features are left to remain the same as those of the control experiment. In the case of ten days, the hurricane frequency does not change much on the day side but decreases significantly on the night side (Fig.~5a). In the case of 20 days, there are nearly no evidence of a hurricane on the night side and the number of hurricanes on the day side also decreases substantially (Fig.~5b). For the case of 40 days, the hurricane only forms at regions very close to the substellar point (Fig.~5c). This trend as a function of rotation period is mainly attributed to three factors: the direct weakness of planetary vorticity, the reduction of relative vorticity due to the fact that the atmosphere becomes so steady that waves and disturbances become less active, and the decrease of relative humidity on the night side (with an increase on the day side) due to the strengthening of the thermal-driven global Walker circulation (Fig.~S2).

When the surface temperature is decreased, it is harder for a hurricane to form. As the maximum surface temperature is set to 308 K and the night-side surface temperature is set to 275 K (Fig.~6a), fewer hurricanes form in the vicinity of the substellar point and no hurricane on the night side (Fig.~5d). When the maximum surface temperature is set to 301 K and the night-side surface temperature is set to 268 K(Fig.~6b), there are only two hurricane events during the integration of four Earth years (Fig.~5e), which is consistent with the GPI prediction in Bin et al. (\cite{Bin}). The temperature of 301 K is close to the tropical surface temperatures on Earth. This suggests that hurricane formation on tidally locked planets requires a warmer surface due to their slower rotation rates and stronger wind shears. For planets with both slow rotation and low temperature, no hurricane can form (Fig.~5f). The decreasing trend of hurricane formation as a function of reduced surface temperature is due to two main processes: the relative humidity and potential intensity decrease because of the cooler surface and weaker upwelling and convection, and the vertical wind shear becomes much stronger due to the enhanced temperature gradients between the day and night sides (Fig.~S3). Moreover, when the night-side surface temperature is low, air convergence from the night side to the day side brings cool air rather than warm air into the substellar region, suppressing hurricane formation there.

\subsection{Effect of bulk atmospheric composition}
Atmospheric compositions on terrestrial exoplanets are as yet unknown. Here, we carry out a preliminary investigation of how atmospheric molecular weight influences hurricane formation under a uniform surface temperature of 301 K. When the background atmosphere is set to H$_2$ or He, there is no hurricane, in contrast to the experiments of N$_2$, O$_2$, and CO$_2$ (Fig.~8), although the GPI value is comparable to or even larger than that shown above. This is due to the fact that the condensate–H$_2$O is heavier than H$_2$ and He, so that any disturbance that brings water vapor upward will cause the density of a moist parcel to be larger than its surrounding environment, similar to the conditions in Saturn’s atmosphere (Guillot \cite{Guillot}, Li \& Ingersoll \cite{LiC}, Leconte et al. \cite{Leconteb}). This process induces a negative buoyancy and stabilizes the atmosphere against convection. It can simply be understood by using the ideal gas equation, $p=\rho R_d T_v$ , and the virtual temperature ($T_v$),
\begin{center}
\begin{equation}
  \centering
 T_v=\frac{p}{p+\left(\epsilon-1\right)e}T
,\end{equation}  
\end{center}

\noindent
where $p$ is total air pressure, $e$ is the partial pressure of the condensate, $\rho$ is air density, $R_d$ is the gas constant of dry air, $\epsilon$ is the molecular weight ratio of water vapor to the dry air, and $T$ is the air temperature. For a H$_2$-dominated (or He-dominated) atmosphere, $\epsilon$ is equal to 9 (or 4.5), so that $T_v$ is smaller than $T$. Therefore, a moist parcel is heavier than a dry parcel under the same $p$ and $T$, and moist convection is inhibited, which is opposite to the conditions on Earth. Moreover, in the experiments with N$_2$, O$_2$, and CO$_2$, a clear trend is revealed, namely, that the size of the hurricane decreases as the mean molecular weight is increased. This is due to the fact that the atmospheric scale height is inversely proportional to the mean molecular weight and, subsequently, the Rossby deformation radius (see the last paragraph of Section 2 above) becomes smaller.

In the experiments of  Fig.~8, the value of Rossby deformation radius is $\approx$500-1500 km, comparable to the hurricane size. However, the Rossby deformation radius is strongly latitude-dependent because $f$ is equal to $2\Omega sin(\varphi)$, where  $\Omega$ is the rotation rate and $\varphi$ is the latitude, but the hurricane size in the experiments does not exhibit the same dependency. A better scaling was not found because of the nonlinear dynamics of hurricane and the complex interactions between the hurricane and diabatic heating, environmental relative humidity, mesoscale convective system, and other features (Emanuel \cite{Emanuelb}; Merlis \& Held \cite{Merlis19}). Under more  realistic conditions, such as on Earth and the simulations in Sections 3.1 and 3.2, the Coriolis effect is not constant between latitudes and there are strong interactions between hurricanes and mean circulation, so that the Rossby deformation radius is not a good scale for hurricane size (e.g., Chavas et al. \cite{Chavas16}).

\section{Conclusions and discussions}

We find that hurricanes can form on tidally locked planets especially for those orbiting near the inner edge of the habitable zone of late M dwarfs. For planets in the middle range of the habitable zone, hurricanes are relatively fewer. Storm theories of Earth and Saturn can be used to understand the hurricane formation on tidally locked planets. Hurricanes can enhance the ocean mixing and oceanic heat transport from warmer to cooler regions in both horizontal and vertical directions. Hurricanes can also influence the transmission spectra of tidally locked planets. For instance, if a hurricane moves to the terminator, water vapor concentration would increase  (Fig.~S4), which can influence the transmission signals. Unfortunately, present-day telescopes are not capable of observing this feature (Morley et al. \cite{Morley}, de Wit et al. \cite{de Wit}) mainly due to the small-scale height of the atmosphere and the relatively small size of the hurricane compared to the planetary radius. Differentiating them requires the large space telescopes or ground-based extremely large telescopes of the future.

Furthermore, future studies require the use of AGCMs coupled to a slab ocean or fully coupled atmosphere–ocean models. The result of a test of the atmosphere coupled to a slab ocean is shown in Fig.~9. We can still find hurricanes in the experiment, whereas more results for different rotation periods, different stellar fluxes, and different CO$_2$ concentrations will be presented in a separate paper in the near future. Moreover, future works are required to examine how continents influence the results of such studies. Hurricanes always decay quickly when they move over land because of the dramatic reduction in evaporation and the increase in surface roughness. Global climate models with more realistic cloud schemes and regional cloud-resolving models with more accurate radiation transfer are required to simulate the hurricane genesis, especially for those who have quite different atmospheric compositions or air masses from Earth. One another weakness in this study is the convection scheme that was developed based on the knowledge of convection on Earth. Future studies using high-resolution models with explicit convection (e.g., Sergeev et al. \cite{Sergeev}) are required. Moreover, convective self-aggregation (such as Bony et al. \cite{Bony}, Pendergrass et al. \cite{Pendergrass}, Wing et al. \cite{Wing}) may have occurred in our simulations, particularly in the 310--315~K experiment.  A future work is required to analyze this feature.

Recently, using Earth-based metrics for hurricane genesis, Komacek et al. (\cite{Komacek}) found that
hurricane genesis is most favorable on tidally locked terrestrial exoplanets with intermediate rotation periods of about 8–10 days in the habitable zones of late-type M dwarf stars, and that on slowly rotating planets hurricane generis is unfavorable. The latter is consistent with Bin et al. (\cite{Bin}) and our results. Future simulations using hurricane-resolved models are required to verify the intermediate rotation periods conclusion shown in Komacek et al. (\cite{Komacek}).

\begin{acknowledgements}
     We thank the National Center for Atmospheric Research (NCAR) groups for developing the model CAM4 and making it available to public. We are grateful to the discussions with Hao Fu, Gan Zhang, Cheng Li, Weixin Xu, Yongyun Hu, Zhiyong Meng, Dorian S. Abbot, Thaddeus D. Komacek, and Fengyi Xie.
\end{acknowledgements}

\begin{appendix} 

\newpage

\section{Supplementary Figures} 

This section includes Figs. S1 to S4.

\renewcommand{\thefigure}{S\arabic{figure}}
\setcounter{figure}{0}
\section{Supplementary Video} 

A video for the evolution of hurricanes on tidally locked planets is included as part of this study. There are two panels in the video, with wind speed at 850~hPa (m\,s$^{-1}$) and surface air pressure (hPa). The drift of hurricanes is mainly due to environmental winds. 

\newpage

\begin{figure*}
\centering
\setlength{\abovecaptionskip}{0.1cm}
\includegraphics[width=0.8\textwidth]{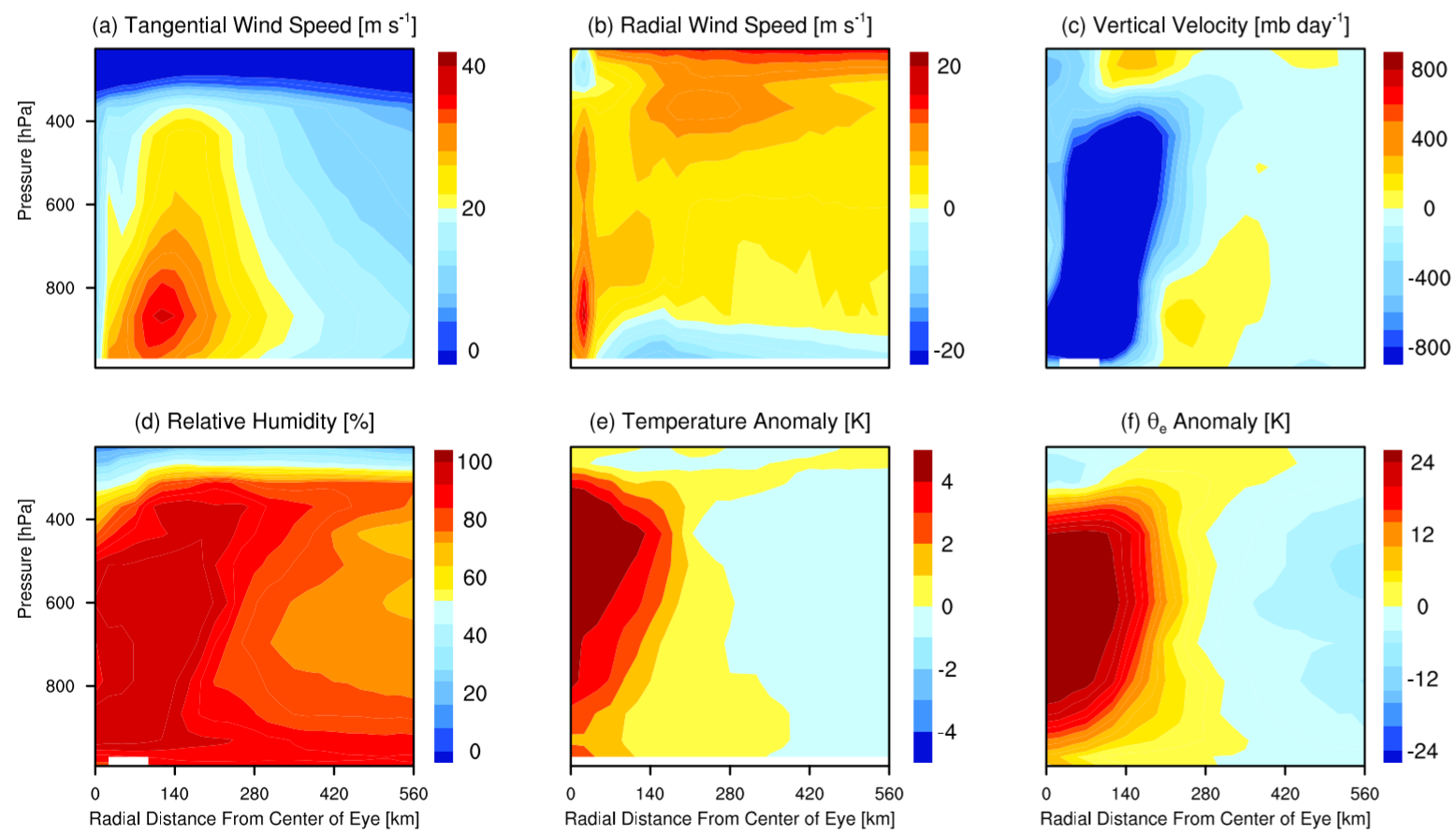}
\caption{Same as Fig.~2 but for a typical, mature hurricane on the night side. The size is about 1/3 of that on the day side.}
\label{figS1}
\end{figure*}

\begin{figure*}
\centering
\setlength{\abovecaptionskip}{0.1cm}
\includegraphics[width=0.8\textwidth]{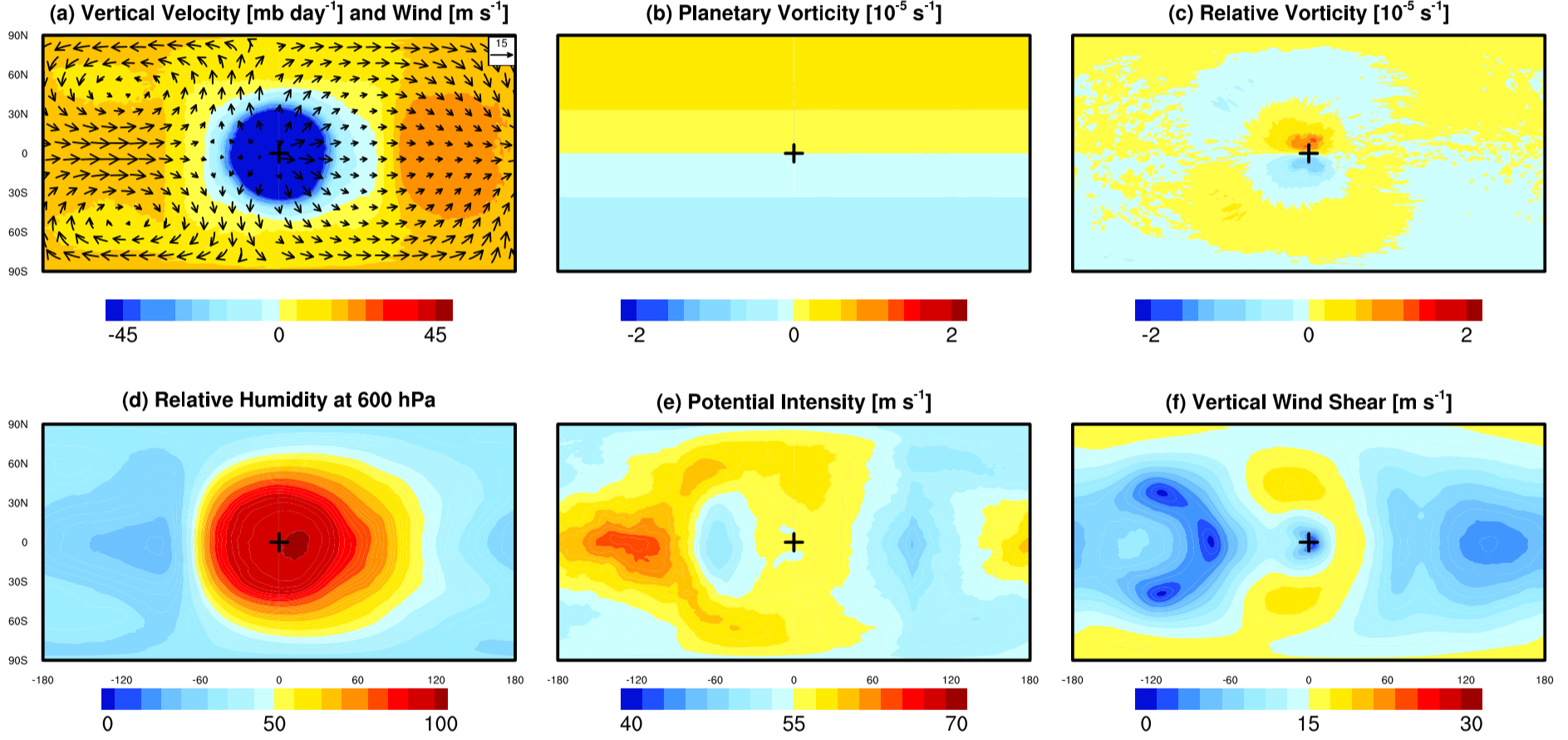}
\caption{Environmental conditions for hurricane formation in the experiment of the rotation period of 40 Earth days and surface temperature shown in Fig.~3a. (a): long-term mean vertical velocity at 500 hPa (shading) and winds at 300 hPa (vector); (b): planetary vorticity; (c): relative vorticity at 850 hPa; (d): relative humidity at 600 hPa; (e): potential intensity; (f): vertical wind shear between 300 and 850 hPa.}
\label{figS2}
\end{figure*}

\begin{figure*}
\centering
\setlength{\abovecaptionskip}{0.1cm}
\includegraphics[width=0.8\textwidth]{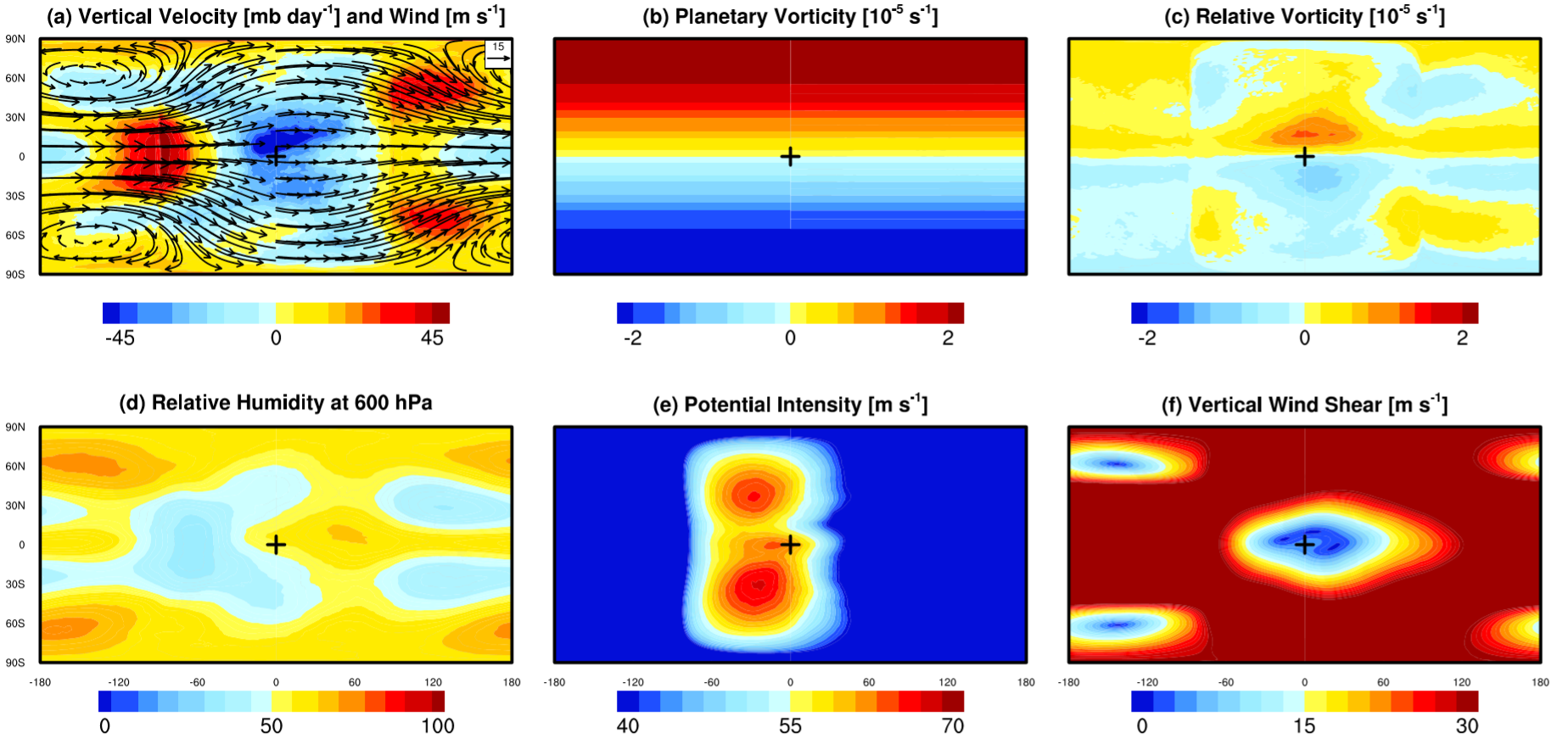}
\caption{Environmental conditions for hurricane formation, same as Fig.~S2 but for the experiment of the rotation period of six Earth days and surface temperature, as shown in Fig.~6a.}
\label{figS3}
\end{figure*}

\begin{figure*}
\centering
\setlength{\abovecaptionskip}{0.1cm}
\includegraphics[width=0.8\textwidth]{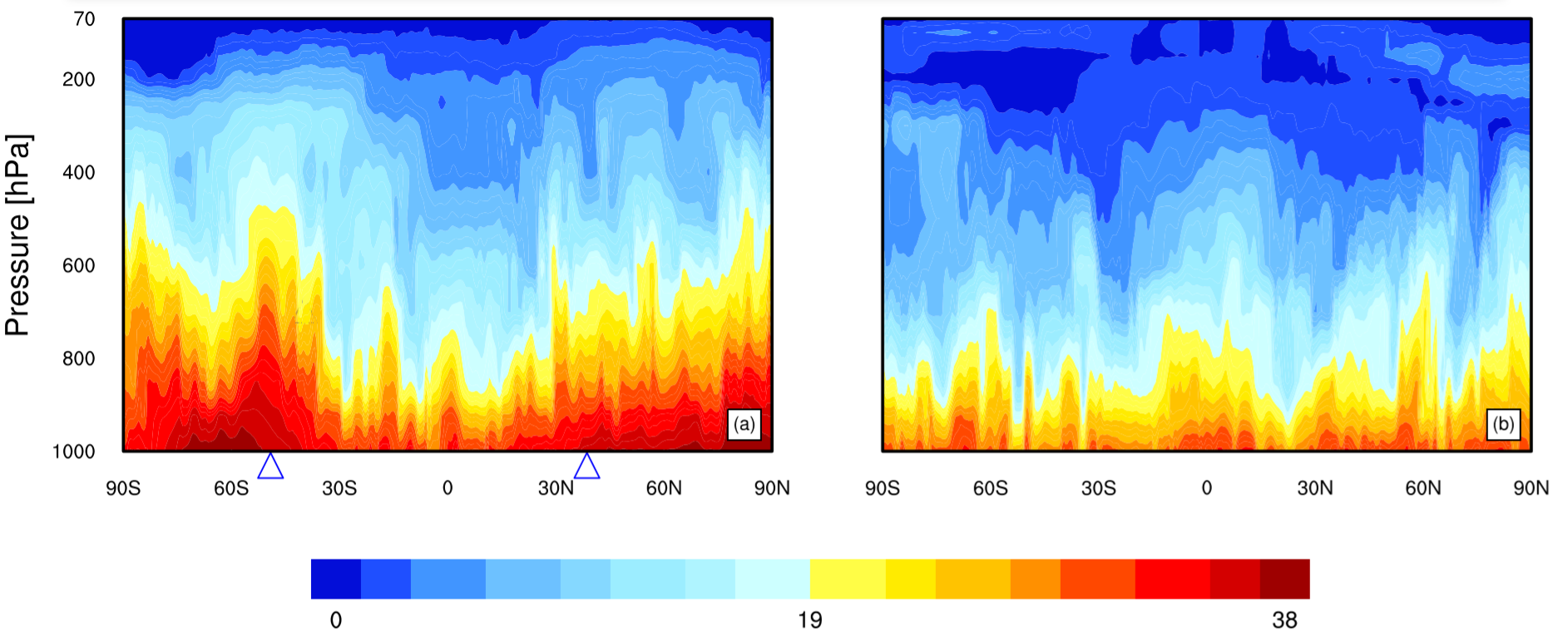}
\caption{Effect of hurricane on water vapor concentration at the west terminator in the control experiment. (a): Specific humidity (g kg$^{-1}$) when two hurricanes are approaching the terminator. (b): Same as (a) but when there is no hurricane. The blue triangle indicates the latitude of the hurricane center.}
\label{figS4}
\end{figure*}

\end{appendix}
\end{document}